\documentclass[conference]{IEEEtran}
\IEEEoverridecommandlockouts

\usepackage{fancyhdr} 
\usepackage{amsmath,epsfig}

\usepackage{amssymb}
\usepackage{bbding}

\usepackage{color}
\usepackage{cite}
\usepackage{amsmath,amssymb,amsfonts}
\usepackage{algorithmic}
\usepackage{graphicx}
\usepackage{textcomp}
\usepackage{xcolor}

\def\BibTeX{{\rm B\kern-.05em{\sc i\kern-.025em b}\kern-.08em
    T\kern-.1667em\lower.7ex\hbox{E}\kern-.125emX}}

\begin{document}

\title{CPNet: Exploiting CLIP-based Attention Condenser and Probability Map Guidance for High-fidelity Talking Face Generation
}

\author{
\IEEEauthorblockN{Jingning Xu$^{1,2}$, Benlai Tang$^{2*}$, Mingjie Wang$^{3}$, Minghao Li$^{2}$, Meirong Ma$^2$}
\IEEEauthorblockA{$^1$School of Software Engineering, Tongji University, Shanghai, China \\
$^2$Department of AI Technology, Transsion, China \\
$^3$School of Sience, Zhejiang Sci-Tech University, China \\
}
}

\maketitle\thispagestyle{fancy}

\begin{abstract}
Recently, talking face generation has drawn ever-increasing attention from the research community in computer vision due to its arduous challenges and widespread application scenarios, \emph{e.g.} movie animation and virtual anchor. Although persevering efforts have been undertaken to enhance the fidelity and lip-sync quality of generated talking face videos, there is still large room for further improvements of synthesis quality and efficiency. Actually, these attempts somewhat ignore the explorations of fine-granularity feature extraction/integration and the consistency between probability distributions of landmarks, thereby recurring the issues of local details blurring and degraded fidelity. To mitigate these dilemmas, in this paper, a novel CLIP-based Attention and Probability Map Guided Network (CPNet) is delicately designed for inferring high-fidelity talking face videos. Specifically, considering the demands of fine-grained feature recalibration, a clip-based attention condenser is exploited to transfer knowledge with rich semantic priors from the prevailing CLIP model. Moreover, to guarantee the consistency in probability space and suppress the landmark ambiguity, we creatively propose the density map of facial landmark as auxiliary supervisory signal to guide the landmark distribution learning of generated frame. Extensive experiments on the widely-used benchmark dataset demonstrate the superiority of our CPNet against state of the arts in terms of image and lip-sync quality. In addition, a cohort of studies are also conducted to ablate the impacts of the individual pivotal components.
\end{abstract}


\section{Introduction}
\label{sec:intro}

Inspired by the roaring success of Convolutional Neural Networks(CNNs) during recent decades, the task of talking face generation has caught increasing attention form the research community in the realm of computer vision~\cite{2020Speech, 2020Talking}. Given any speech signal inputs, talking face generation hammers at synthesizing high-quality and realistic face videos, \emph{i.e.} it converts the speech contents to corresponding visual signals. Thanks to fascinating application scenarios, such as movie animation, virtual computer games and virtual anchor, \emph{etc.}, a series of pioneering attempts have been made with the goal of consistently boosting the synthesis performance and efficiency via sophisticated schemes of representation learning~\cite{2022AnyoneNet, 2022TTS, 2021dubbing, 2021Investigating, 2021Flow}. Albeit the impressive improvements, there is still large room for enhancing image synthesis and lip-sync quality simultaneously~\cite{2022StableFace}.

To realize the expectation of obtaining realistic talking face video, several existing algorithms~\cite{2022AnyoneNet, 2021Multimodal, 2021Investigating} resort to two-stage strategies. They commonly decompose the holistic problem into two sub-stages: \emph{landmark extraction} and \emph{mapping relation learning}. However, the intermediate landmark prediction easily introduces ambiguity into the subsequent generator while suffering form the lack of detailed information (\emph{e.g.} skin texture and background scenarios), thereby resulting in the higher requirements of the recording environment of the data~\cite{2021dubbing}. Collecting such type of data is tedious and arduous in practical applications. Hence, a train of approaches~\cite{2021Investigating, 2022AnyoneNet,2021dubbing, 2019FewShot} refine the generation procedure and probe into the protocol that feeds 2D landmark-equipped reference image into the pipeline. This refinement allows the networks to further enrich texture and background hints. Despite more detailed cues, these approaches are prone to generate stiff and dull expressions since they rely on a single reference image which involves extremely limited information on facial movement and texture details. To delve into the application of auxiliary signals, Yu \emph{et al.}~\cite{2021Multimodal} leverage Canny edge detector to mine geometric hints on hair, clothing and background. However, it is incompetent to endow the model with multi-scale or multi-level property, thereby failing to capture fine-granularity representations.

The inconsistency between landmark inputs and generated faces is another intractable problem. It is worth noting that landmark is incapable of providing lip/teeth details and therefore the inadequate information easily lowers consistency between ground-truth landmarks and predicted mouth regions~\cite{2022StableFace}. To attenuate this issue, studies~\cite{2020Speech, 2022StableFace, wav2lip} propose audio aggregation module and attach it to the primary generation stem. However, the heterogeneity caused by multifarious modals hinders the better aggregation of image and audio features, and requires more sophisticated learning architectures or mechanisms. Although a powerful offline lip-sync discriminator is adopted in Wav2Lip~\cite{wav2lip} as a type of auxiliary signals to assist in the optimization of generator, speaker-specific models provided by Wav2Lip present issues of blurring faces and inconsistent textures~\cite{lipsync3d}. Additionally, approaches~\cite{2021Multimodal, 2022StableFace} try to strengthen the stability of generation from the perspective of inter-frame smoothness, but the inherited inconsistency problem is overlooked.


The aforementioned drawbacks stimulate our explorations to improve the image fidelity and lip-sync quality while tackling the inconsistency between landmark ground truth and predicted face frame. In this paper, we propose a Clip-based attention and Probability map guided Network (CPNet) for high-fidelity talking face generation. In specific, inspired by high-efficiency feature reuse in pattern of dense connections~\cite{2017DenseNet}, we marry our generation backbone with the property of dense feature reuse to mine multi-scale and multi-level semantics.  
Furthermore, to enrich fine-granularity representations and absorb higher-level priors, we delicately exploit a clip-based~\cite{2021CLIP} attention condenser to transfer knowledge with sufficient multi-modal semantic cues from CLIP and recalibrate the intermediate feature channels of our generator. Last but by no means least, motivated by the prevalence of density map in crowd counting task~\cite{2022Density, wang2022stnet, wang2023dynamic}, we novelly present probability map of landmarks for constraining the consistency between the generated face frame and the landmark groundtruths in probability space instead of inchoate pixel-wise Euclidean distance. The introduction of probability map is also beneficial for imposing landmark distribution learning on our CPNet. 

In a nutshell, our main contributions are fourfold.

\begin{itemize}
\item {\bf In use:} In this paper, we propose a novel Clip-based attention and Probability map guided Network (CPNet) for effective and high-fidelity talking face generation.
\item {\bf Fine granularity:} A \emph{clip-based attention mechanism} is delicately designed to extract fine-granularity representations by transferring priors with rich semantic cues from the prevailing CLIP paradigm.
\item {\bf Consistency:} We creatively propose a type of constraints, \emph{probability map}, to guarantee the consistency or smoothness between the generated frames and the probability distributions of the ground-truth landmarks.
\item {\bf Performance:} Extensive experiments and ablation studies demonstrate the superiority of our proposed framework in terms of the fidelity and lip-sync quality.
\end{itemize}

\section{Related Work}

\label{sec:rw}

\subsection{Landmark-to-Face Video Generation}

Benefiting from the booming of GAN-based video synthesis, the majority of existing studies attempt to regress the video frames from facial landmark inputs for talking face generation via GAN-based methods~\cite{2022AnyoneNet, 2021Multimodal, 2021Investigating, 2018obamanet}. 
The work~\cite{2017obama} carefully designs a rendering framework based on texture searching and selection to synthesize Obama videos with about 17 hours footage. U-Net and implicit condition are adopted in the study~\cite{2018obamanet} to draw an outline of mouth on the cropped input scenes. This strategy makes the rendered mouth region more seamless with the top half part in the target video. These approaches attempt to regress mouth regions associated with speech content. In contrast, several studies employ reference images to assist in the generation under different scenarios~\cite{2022AnyoneNet, 2021Multimodal, 2021Investigating}. Wherein, Kesim \emph{et al.}~\cite{2021Investigating} and AnyoneNet~\cite{2022AnyoneNet} utilize a reference image from a specified scenario to provide scenario details and speaker texture characterization for fine-grained generation. Yu \emph{et al.}~\cite{2021Multimodal} adopt a pix2pixHD to generate the initial frame, and treat foregoing generated images as the input for subsequent generation to guide the learning of robust representations.

\subsection{CLIP-based Knowledge Transfer}

Recently, the advent of large-scale visual-language pre-training models reveals their powerful capability of representing semantically rich and high-level visual concepts through natural language supervision. Wherein, the most representative and popular framework is Contrastive Language-Image Pre-training (CLIP)~\cite{2021CLIP} and a vast number of clip-based methods are in vogue in computer vision as it contains abundant multi-modal knowledge for a variety of down-stream visual tasks. In the field of image generation, several models~\cite{clip1, clip2} make attempts to transfer semantically-rich priors from clip, and attains impressive results. Studies~\cite{clip4} use pre-trained clip to model relationship between image and input text, and illustrate that the latent space of clip is capable of semantically modifying images by moving along the dimension of text embedding. The work~\cite{clip2} endeavors to introduce a CLIP latent residual mapper that is trained for a specific text. By doing so, the hybrid collaboration of strong generation ability of StyleGAN~\cite{karras2019style} and the extraction of extraordinary visual concepts is achieved. Albeit the great potentials of CLIP, how to transfer priors from CLIP for capturing fine-granularity representations in the realm of image generation is still a burning problem.


\begin{figure*}[!t]

\centering
\includegraphics[width=18cm]{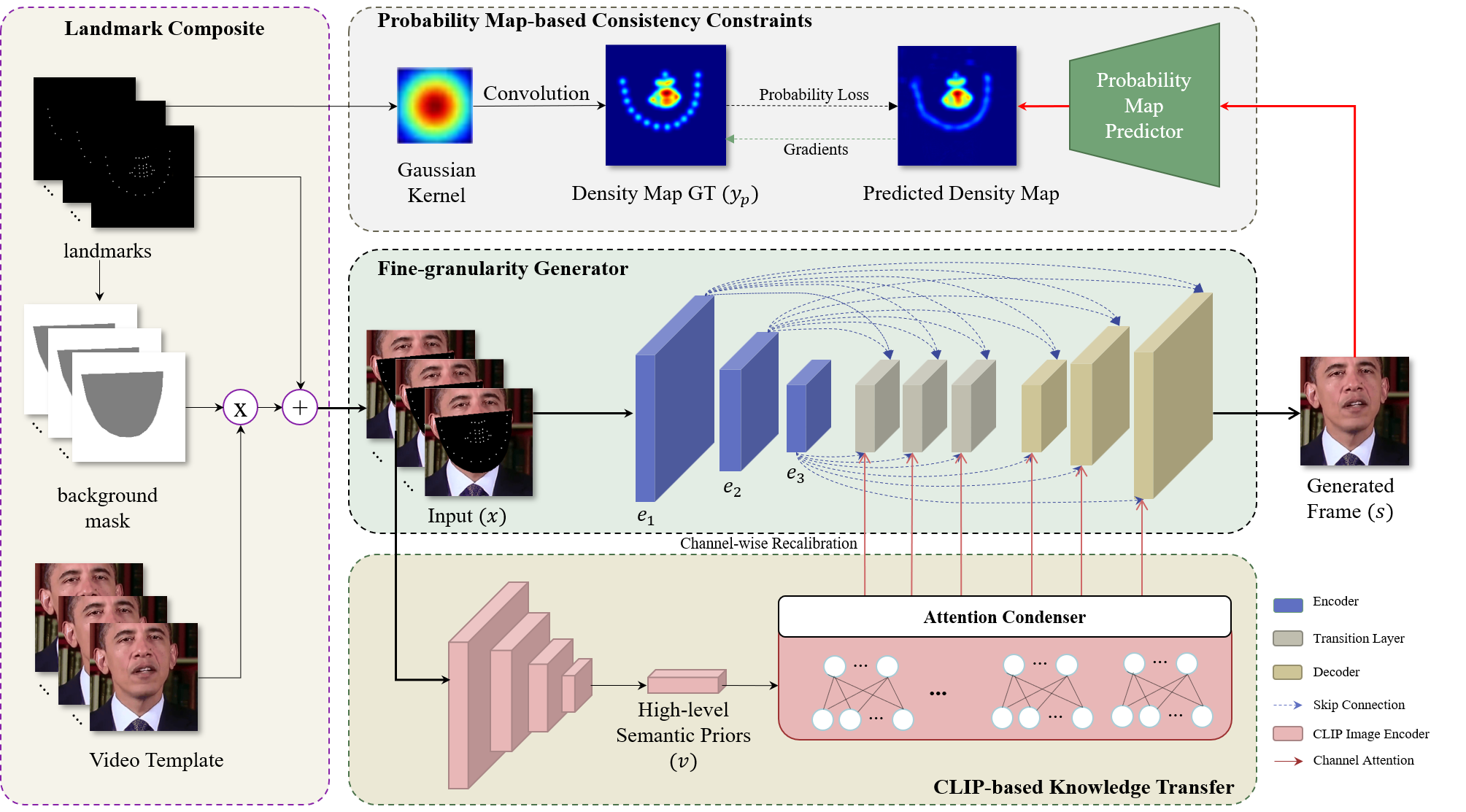}
\caption{The overview of the proposed CPNet towards exploiting CLIP-based attention condenser and probability map guidance for high-fidelity talking face generation, which is composed of three modules that includes densely-connected Generator, CLIP-based Knowledge Transfer and Probability Map Predictor.
}
\label{generator}
\end{figure*}

\section{Methodology}
\subsection{Densely-connected Generation Backbone}
Following existing state-of-the-art image translation methods~\cite{2022AnyoneNet, 2018obamanet, 2021dubbing}, the commonly-used pix2pix is adopted as our backbone to regress the realistic pixel-level images. Analogous to the model~\cite{2021dubbing}, we feed seven consecutive frames involving the current one frame and three prior frames into our backbone rather than inputting current single scene. Treating a batch of sequential frames as input contributes to model inter-frame relationships. To expand the diversities of receptive fields and semantic levels for intermediate features, the pattern of dense connection~\cite{2017DenseNet} is employed in our backbone to reuse features from all preceding transition layers. Apart from the re presentation richness, another merit of dense connection is that it can alleviate gradient vanishing and speed up the convergence of the model. It implies that for any layer $x^l$ in transition layer or decoder, the input is modified as: 
\begin{equation}
x^l = x^l + \sum_{i=1}^{3}Pool(H^l_i(e_i))
\end{equation}
where $e_i$ is the output of the $i_th$ layer of encoder, $H^l_i$ is a convolution operation with kernel size 1x1 and $Pool(\cdot)$ is an adaptive pooling operation to perform multi-scale feature map size alignment.


\subsection{CLIP-based Attention Condenser}
As described in Section~\ref{sec:rw}, the large-scale pre-trained CLIP is widely used in various image generation tasks thanks to its rich semantic priors through multi-modal contrastive learning. Here we reload the weights of ViT part in CLIP and take the facial landmark image of the current frame as input of clip-based knowledge transfer module. And then ViT stem outputs a one-dimensional vector as the high-level semantic priors. Furthermore, this vector is passed through a SENet~\cite{SENet}-like sub-network to transfer the semantic priors into the channel-wise attention weights. The key difference is that the recalibration weights of our attention condenser are derived from the linear transformation of CLIP latent embedding with sufficient semantic information instead of self-channel features. On top of the adaptively-learned weights, intermediate representations of the generator could be further enhanced by absorbing semantic cues provided by CLIP. And then for each layer $x^l$ in transition layer or decoder, we opt to employ a simple gating mechanism with a sigmoid activation: 
\begin{equation}
x^l = F_{scale}(x^l, \sigma(Wv)),
\end{equation}
where $F_{scale}(\cdot)$ indicates the channel-wise multiplication between the scalar and the feature map, whereas $v$ denotes the CLIP latent codes, $W$ denotes the weights of linear transformation and $\sigma(\cdot)$ represents the sigmoid function.

\subsection{Landmark Probability Map}
To guarantee the consistency between the generated talking face and the corresponding ground-truth landmark image in the probability space, we propose a scheme to constraints the predicted frame via auxiliary probability map predictor and density map-based loss function. Probability map is produced by convolving the original facial landmark dot image with a heuristically-defined Gaussian kernel. The results empirically show that when the size of Gaussian kernel is set as 25x25 and sigma is configured as 5 in our task, the probability map-based constraint makes the best positive impacts on the holistic learning procedure of our model.

The probability map predictor is built for the generation of predicted density map of landmarks. The predictor adopts pix2pix network architecture with lightweight parameters and is placed at the end of pipeline.
Furthermore, the predictor is not expected to have too much robustness. In other words, the prediction module should be sensitive and discriminative to small changes in the talking face image. Following the design idea of hinge loss, the training objective of this prediction module is set as: 
\begin{equation}
l_{dmp} = ||P(I) - y_{p}||_2 - \lambda ||P(I') - P(I)||_2,
\end{equation}
where $P$ denotes the prediction module and $y_{p}$ denotes the target probability map and $I$ is the real image, whereas $I'$ represents the image generated by the generator. $\lambda$ is a adjustment hyperparameter to control the effects of the prediction module.

\subsection{Objective Function}
The ultimate training objective is to supervise the optimization of our proposed framework for regressing high-fidelity talking face video with high lip-sync quality. To simplify the subsequent explanations, the notation is defined here. We denote the sequence of consecutive frames input to the generator as $x$, whereas the corresponding ground truth frame is $y$ and $s$ indicates the generated frame.

Following LSGAN loss~\cite{2017Least}, we first implement the adversarial loss $L_{adv}$ as: 
\begin{equation}
L_{adv}=\mathbb{E}_{x,y}[(D(x, y)-1)^2]+\mathbb{E}_{x,s}[D(x, s)^2],
\end{equation}
where $\mathbb{E}[\cdot]$ denotes expectation values and $D$ means the discriminator network.

Meanwhile, following the configuration in ~\cite{2016Perceptual}, the perceptual reconstruction loss $L_r$ provides the supervisory signal to impel the generated frames to be close to the ground truths. By introducing perceptual loss term, the quality of predicted images can be improved based on the differences between high-level image representations extracted from pre-trained convolutional neural networks, such as VGGNet~\cite{simonyan2014very}. The perceptual perceptual loss $l_r$ is defined as: 
\begin{equation}
L_r = \frac{1}{l}\sum_l||\phi^l(s) - \phi^l(y)||_1,
\end{equation}
where $\phi^l$ is the $l^{th}$ layer of the pre-trained VGGNet.

Moreover, the video loss $L_t$ is implemented by sequence discriminator $D_t$ to facilitate the video quality, especially for spatio-temporal coherence. Similar to $L_{adv}$, $L_{t}$ is defined as: 
\begin{equation}
L_{t}=\mathbb{E}_{x^t,y^t}[(D_t(x^t, y^t)-1)^2]+\mathbb{E}_{x^t,s^t}[D_t(x^t, s^t)^2],
\end{equation}
where $D_t$ denotes the discriminator network, $x^t, y^t, s^t$ are the corresponding input, ground truth and the generated image, respectively, for consecutive several frames.

To enhance the consistency in probability space, a pivotal probability map loss term is designed here to shorten the distance between the generated image and ground-truth landmark distribution. We denote the probability map predictor as $P$ and the probability map loss $l_{p}$ is written as: 
\begin{equation}
L_{p} = ||P(s) - P(y)||_2
\end{equation}
In summary, the entire loss function of our CPNet is formulated as: 
\begin{equation}
L = \lambda_{adv}L_{adv} + \lambda_{r}L_{r} + \lambda_{t}L_{t} + \lambda_{p}L_{p},
\end{equation}
where $\lambda_{adv}, \lambda_{r}, \lambda_{t}, \lambda_{p}$ are hyperparameters which are utilized to adjust the influences of multifarious loss terms.

\section{Experiments}

\subsection{Datasets and Implementation Details}

\noindent\textbf{Datasets.} Our CPNet is verified on the prevailing ObamaSet benchmark~\cite{2017obama} which is widely chosen to evaluate lip-sync methods. This dataset includes the total of 17 hours  videos at 29.97 fps with a resolution of 1280$\times$720. This dataset contains hundreds of scenarios from Obama's weekly speeches with a wide range of clothing types, lighting conditions and backgrounds. We randomly select 200 videos in total of 3 hours from the original dataset. Wherein, 90$\%$ of the samples are used as the training set while the rest of 10\% is treated as the testing set to testify the performance of talking face generation during the inference phase.

\noindent\textbf{Evaluation Metrics.}
In our experiments, to evaluate the quality of the synthesized frames, common reconstruction metrics like the peak signal-to-noise ratio (PSNR) and the structural similarity (SSIM) metrics~\cite{PSNRSSIM} are used to reflect the visual quality. And the Fréchet Video Distance (FVD) is used to evaluate the spatio-temporal consistency of the generated videos~\cite{2018FVD}.
Besides, “LSE-D" and “LSE-C", proposed in~\cite{wav2lip}, is adopted to evaluate the lip-sync quality of generated videos. The former calculates the distance between the lip and audio representations and the latter metric the average confidence score between the speech and lip movements.

\noindent\textbf{Implementation Details.}
The weights for the loss terms are empirically fixed as: $\lambda_{adv} = 1, \lambda_{r} = 5, \lambda_{t} = 1, \lambda_{p} = 0.1$. We adopt Adam optimizer with the learning rate of 0.0001 and the momentum of 0.5 to optimize our CPNet for around 100k iterations. During the training phase, frames with the size of 224$\times$224 are centrally cropped around the center of facial region. 





\begin{figure*}[!t]

\centering
\includegraphics[width=17.5cm]{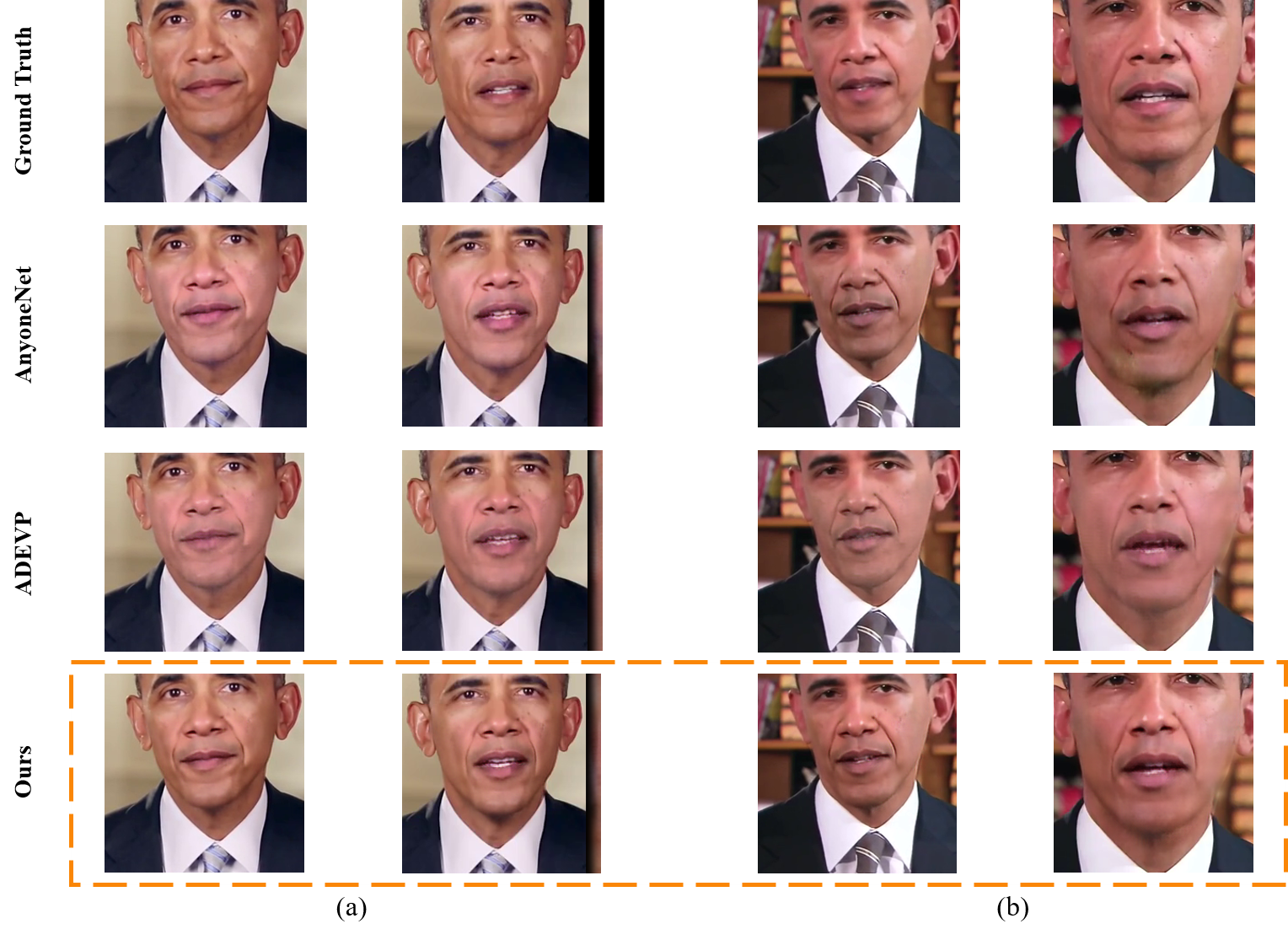}

\vspace{-0.4cm}
\caption{The visualization comparisons among baselines and our CPNet on two scenarios ((a) and (b)) from ObamaSet dataset. It can be observed that our method produces more realistic results than AnyoneNet~\cite{2022AnyoneNet} and ADEVP~\cite{ji2021audio}, especially for the generations of skin and lip textures. 
}
\label{comparison}
\vspace{-0.4cm}
\end{figure*}

\subsection{Comparison with State of the Arts}

We reimplement the ``landmark to photo-realistic image'' module in AnyoneNet~\cite{2022AnyoneNet} and ``Edge-to-Video'' module in ADEVP~\cite{ji2021audio} as baselines, whose objective is essentially the same as ours. The quantitative results are demonstrated in Table~\ref{tab:comparison} while the visualization examples are also provided in Figure 2. Both quantitative results and visualization examples demonstrate the superiority of our proposed framework. It can be concluded that our proposed approach is better than existing methods in all aspects of image quality, video consistency, and lip-sync quality.

\begin{table}[htbp]


\vspace{-0.4cm}
\caption{Comparisons with baseline models on ObamaSet. The proposed method consistently surpasses the existing approaches on all metrics. ``$\uparrow$'' means the greater the metric, the better the generation, and ``$\downarrow$'' is the opposite.}
\centering
\resizebox{\linewidth}{!}{
\begin{tabular}{c|c|c|c|c|c}
\hline
Method & SSIM $\uparrow$ & PSNR $\uparrow$ & FVD $\downarrow$ & LSE-C $\uparrow$ & LSE-D $\downarrow$ \\ \hline
AnyoneNet\cite{2022AnyoneNet} &  0.959 & 34.0 & 0.146 & 0.236 & 14.98 \\ 
ADEVP\cite{ji2021audio}  & 0.963 & 34.6 & 0.113 & 0.299 & 14.63 \\ 
Our approach &  \textbf{0.973} & \textbf{35.6} & \textbf{0.060} & \textbf{0.401} & \textbf{13.66} \\ \hline
\end{tabular}
}
\label{tab:comparison}

\end{table}

\vspace{-0.3cm}
\subsection{Ablation Study}

\subsubsection{The Impacts of Individual Components}

To better understand each component in our model and verify the effects of them, we carry out various ablation studies on ObamaSet benchmark (see Table~\ref{tab:ablation}). The results demonstrates that densely-connected generator (\emph{I}) has an impressive contribution to visual quality. Thanks to the introduction of rich semantic information, the image quality and fidelity of the generated results have been greatly improved by exploiting the attention condenser through CLIP latent embedding (\emph{II}). And the probability map constraints of landmark (\emph{III}) also have positive impacts on the lip-sync metrics.


\begin{table}[htbp]

\vspace{-0.4cm}
\caption{Ablation studies for proposed individual components: densely-connected generator (\emph{I}), CLIP-based attention condenser (\emph{II}) and probability map constraints (\emph{III}).}
\centering

\resizebox{\linewidth}{!}{
\begin{tabular}{c c c |c|c|c|c|c}
\hline
\emph{I} & \emph{II} & \emph{III} & SSIM $\uparrow$ & PSNR $\uparrow$ & FVD $\downarrow$ & LSE-C $\uparrow$ & LSE-D $\downarrow$ \\ \hline
 &  &  & 0.950 & 32.9 & 0.157 & 0.197 & 15.33 \\ 
\Checkmark &  &  &  0.961 & 34.1 & 0.108 & 0.23 & 14.85 \\ 
\Checkmark & \Checkmark &  & 0.971 & 35.4 & 0.065 & 0.312 & 14.21 \\ 
\Checkmark & \Checkmark & \Checkmark & \textbf{0.973} & \textbf{35.6} & \textbf{0.060} & \textbf{0.401} & \textbf{13.66} \\ \hline
\end{tabular}
}
\label{tab:ablation}
\end{table}

\subsubsection{How do Loss Terms Influence the Performance}


To further investigate the effects of all loss terms in our objective function on the performance, we conduct extensive experiments to ablate all loss terms in Table~\ref{tab:loss}. It can be observed that using only standard GAN loss of $l_{adv}$ to guide the generation is rather terrible. Additionally, $l_r$ is designed to match the features extracted by the VGG network and allows for alleviating the issue of image blurring and bringing remarkable improvements in image quality metrics. Specifically, compared to the results without $l_t$, the FVD scores are significantly increased and the quality of the generated image is substantially improved with $l_t$. Finally, the introduction of $l_p$ significantly improves the quality of the generated lip-sync results.

\begin{table}[htbp]


\vspace{-0.4cm}
\caption{The impacts of multifarious loss terms on model performance.}

\centering

\resizebox{\linewidth}{!}{
\begin{tabular}{c|c|c|c|c|c}

\hline
Loss Functions & SSIM $\uparrow$ & PSNR $\uparrow$ & FVD $\downarrow$ & LSE-C $\uparrow$ & LSE-D $\downarrow$ \\ \hline
$L_{adv}$  &  0.966 & 34.7 & 0.151 & 0.252 & 14.69 \\ 
$L_{adv} + L_r$  &  0.970 & 35.5 & 0.129 & 0.288 & 14.51 \\ 
$L_{adv} + L_r + L_t$ & 0.971 & 35.4 & 0.065 & 0.312 & 14.21 \\ 
$L_{adv} + L_r + L_t + L_{p}$ & \textbf{0.973} & \textbf{35.6} & \textbf{0.060} & \textbf{0.401} & \textbf{13.66} \\ \hline
\end{tabular}
}
\label{tab:loss}

\end{table}

\begin{figure}[!t]

\centering
\includegraphics[width=8.5cm]{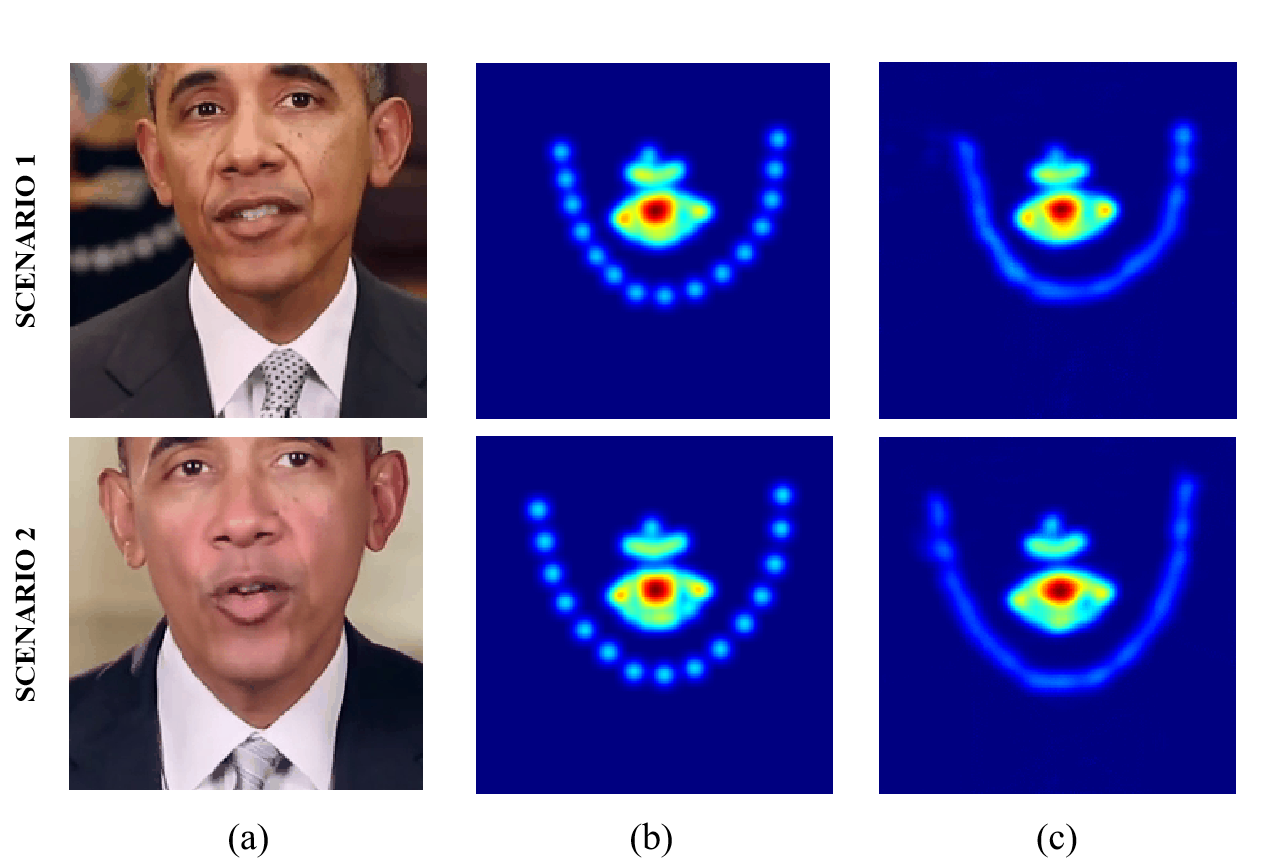}
\caption{Probability map prediction results on different scenarios. (a) the generated frames; (b) the ground truths of probability map; (c) the predicted probability maps by our model.}
\label{generator}

\end{figure}

As for the hyperparameters in our loss function, multiple sets of experiments are carried out and we expirically fix the final weights. Taking $L_p$ as an example, we set weight as 0.05, 0.1, 0.5 and 1.0 for four sets of experiments and obtained the relevant results, see Table~\ref{tab:weight}. As shown in Table~\ref{tab:weight}, the best performance is obtained when the value of $L_p$ is 0.1. We hence fix the $L_p$ as value of 0.1. The controlling weights for remaining loss terms are also determined in this way.

\begin{table}[htbp]

\vspace{-0.4cm}
\caption{The effects of hyperparameters in objective function.}
\centering

\resizebox{\linewidth}{!}{
\begin{tabular}{c|c|c|c|c|c}

\hline
Hyperparameter $\lambda_p$ & SSIM $\uparrow$ & PSNR $\uparrow$ & FVD $\downarrow$ & LSE-C $\uparrow$ & LSE-D $\downarrow$ \\ \hline
1.0  &  0.965 & 34.6 & 0.063 & \textbf{0.419} & \textbf{13.58} \\ 
0.5  &  0.967 & 34.8 & 0.066 & 0.415 & 13.61 \\ 
0.1 & \textbf{0.973} & \textbf{35.6} & \textbf{0.060} & 0.401 & 13.66 \\ 
0.05 & 0.972 & 35.4 & 0.061 & 0.342 & 14.07 \\ \hline
\end{tabular}
}

\label{tab:weight}
\end{table}

\section{Conclusion}

In this paper, we exploit clip-based attention condenser and probability map guidance for high-fidelity and lip-sync talking face generation. Specifically, a densely-connected generation backbone is proposed to mine fine-granularity representations with diverse scales and levels. Then, an attention condenser is delicately designed to transfer priors from pre-trained CLIP models to enrich the multi-modal semantic cues. Finally, we devise a new probability map constraint of landmark to guarantee the consistency between generated images and groundtruths in probability space of landmark. Extensive experiments and ablation studies on ObamaSet dataset illustrate the effectiveness and superiority of our proposed CPNet.

\section*{Acknowledgment}

We would like to express our sincere gratitude to our colleagues of engineering services team who provided essential computational resources during experiments. Additionally, we are also grateful for the support provided by the Science Foundation of Zhejiang Sci-Tech University (ZSTU) under Grant No. 22062338-Y.

\bibliographystyle{IEEEbib}
\bibliography{icme2023template}

\begin{thebibliography}{10}

\bibitem{2020Speech}
Wentao Wang, Yan Wang, Jianqing Sun, Qingsong Liu, Jiaen Liang, and Teng Li,
\newblock ``Speech driven talking head generation via attentional landmarks
  based representation.,''
\newblock in {\em INTERSPEECH}, 2020.

\bibitem{2020Talking}
Aihua Zheng, Feixia Zhu, Hao Zhu, Mandi Luo, and Ran He,
\newblock ``Talking face generation via learning semantic and temporal
  synchronous landmarks,''
\newblock in {\em ICPR}, 2021.

\bibitem{2022AnyoneNet}
Xinsheng Wang, Qicong Xie, Jihua Zhu, Lei Xie, and Odette Scharenborg,
\newblock ``Anyonenet: Synchronized speech and talking head generation for
  arbitrary persons,''
\newblock {\em IEEE Transactions on Multimedia}, 2022.

\bibitem{2022TTS}
Hyoung-Kyu Song, Sang~Hoon Woo, Junhyeok Lee, Seungmin Yang, Hyunjae Cho,
  Youseong Lee, Dongho Choi, and Kang-wook Kim,
\newblock ``Talking face generation with multilingual tts,''
\newblock in {\em CVPR}, 2022.

\bibitem{2021dubbing}
Tianyi Xie, Liucheng Liao, Cheng Bi, Benlai Tang, Xiang Yin, Jianfei Yang,
  Mingjie Wang, Jiali Yao, Yang Zhang, and Zejun Ma,
\newblock ``Towards realistic visual dubbing with heterogeneous sources,''
\newblock in {\em ACM MM}, 2021.

\bibitem{2021Investigating}
Ege Kesim and Engin Erzin,
\newblock ``Investigating contributions of speech and facial landmarks for
  talking head generation.,''
\newblock in {\em Interspeech}, 2021.

\bibitem{2021Flow}
Z.~Zhang, L.~Li, Y.~Ding, and C.~Fan,
\newblock ``Flow-guided one-shot talking face generation with a high-resolution
  audio-visual dataset,''
\newblock in {\em CVPR}, 2021.

\bibitem{2022StableFace}
Jun Ling, Xu~Tan, Liyang Chen, Runnan Li, Yuchao Zhang, Sheng Zhao, and
  Li~Song,
\newblock ``Stableface: Analyzing and improving motion stability for talking
  face generation,''
\newblock {\em arXiv preprint arXiv:2208.13717}, 2022.

\bibitem{2021Multimodal}
Lingyun Yu, Jun Yu, Mengyan Li, and Qiang Ling,
\newblock ``Multimodal inputs driven talking face generation with
  spatial--temporal dependency,''
\newblock {\em IEEE Transactions on Circuits and Systems for Video Technology},
  vol. 31, no. 1, pp. 203--216, 2020.

\bibitem{2019FewShot}
Egor Zakharov, Aliaksandra Shysheya, Egor Burkov, and Victor Lempitsky,
\newblock ``Few-shot adversarial learning of realistic neural talking head
  models,''
\newblock in {\em ICCV}, 2019.

\bibitem{wav2lip}
KR~Prajwal, Rudrabha Mukhopadhyay, Vinay~P Namboodiri, and CV~Jawahar,
\newblock ``A lip sync expert is all you need for speech to lip generation in
  the wild,''
\newblock in {\em ACM MM}, 2020.

\bibitem{lipsync3d}
Avisek Lahiri, Vivek Kwatra, Christian Frueh, John Lewis, and Chris Bregler,
\newblock ``Lipsync3d: Data-efficient learning of personalized 3d talking faces
  from video using pose and lighting normalization,''
\newblock in {\em CVPR}, 2021.

\bibitem{2017DenseNet}
Gao Huang, Zhuang Liu, Laurens Van Der~Maaten, and Kilian~Q Weinberger,
\newblock ``Densely connected convolutional networks,''
\newblock in {\em CVPR}, 2017.

\bibitem{2021CLIP}
Alec Radford, Jong~Wook Kim, Chris Hallacy, Aditya Ramesh, Gabriel Goh,
  Sandhini Agarwal, Girish Sastry, Amanda Askell, Pamela Mishkin, Jack Clark,
  et~al.,
\newblock ``Learning transferable visual models from natural language
  supervision,''
\newblock in {\em ICML}, 2021.

\bibitem{2022Density}
Jia Wan, Qingzhong Wang, and Antoni~B Chan,
\newblock ``Kernel-based density map generation for dense object counting,''
\newblock {\em IEEE Transactions on Pattern Analysis and Machine Intelligence},
  2020.

\bibitem{wang2022stnet}
Mingjie Wang, Hao Cai, Xianfeng Han, Jun Zhou, and Minglun Gong,
\newblock ``Stnet: Scale tree network with multi-level auxiliator for crowd
  counting,''
\newblock {\em IEEE Transactions on Multimedia}, 2022.

\bibitem{wang2023dynamic}
Mingjie Wang, Hao Cai, Yong Dai, and Minglun Gong,
\newblock ``Dynamic mixture of counter network for location-agnostic crowd
  counting,''
\newblock in {\em Proceedings of the IEEE/CVF Winter Conference on Applications
  of Computer Vision}, 2023, pp. 167--177.

\bibitem{2018obamanet}
Rithesh Kumar, Jose Sotelo, Kundan Kumar, Alexandre de~Br{\'e}bisson, and
  Yoshua Bengio,
\newblock ``Obamanet: Photo-realistic lip-sync from text,''
\newblock {\em arXiv preprint arXiv:1801.01442}, 2017.

\bibitem{2017obama}
Supasorn Suwajanakorn, Steven~M Seitz, and Ira Kemelmacher-Shlizerman,
\newblock ``Synthesizing obama: learning lip sync from audio,''
\newblock {\em ACM Transactions on Graphics (ToG)}, vol. 36, no. 4, pp. 1--13,
  2017.

\bibitem{clip1}
Or~Patashnik, Zongze Wu, Eli Shechtman, Daniel Cohen-Or, and Dani Lischinski,
\newblock ``Styleclip: Text-driven manipulation of stylegan imagery,''
\newblock in {\em CVPR}, 2021.

\bibitem{clip2}
Rinon Gal, Or~Patashnik, Haggai Maron, Amit~H Bermano, Gal Chechik, and Daniel
  Cohen-Or,
\newblock ``Stylegan-nada: Clip-guided domain adaptation of image generators,''
\newblock {\em ACM Transactions on Graphics (TOG)}, vol. 41, no. 4, pp. 1--13,
  2022.

\bibitem{clip4}
Nasir Khalid, Tianhao Xie, Eugene Belilovsky, and Tiberiu Popa,
\newblock ``Clip-mesh: Generating textured meshes from text using pretrained
  image-text models,''
\newblock {\em ACM Transactions on Graphics (TOG).}, vol. 3, 2022.

\bibitem{karras2019style}
Tero Karras, Samuli Laine, and Timo Aila,
\newblock ``A style-based generator architecture for generative adversarial
  networks,''
\newblock in {\em CVPR}, 2019.

\bibitem{SENet}
Jie Hu, Li~Shen, and Gang Sun,
\newblock ``Squeeze-and-excitation networks,''
\newblock in {\em CVPR}, 2018.

\bibitem{2017Least}
X.~Mao, Q.~Li, H.~Xie, Ryk Lau, and S.~P. Smolley,
\newblock ``Least squares generative adversarial networks,''
\newblock in {\em ICCV}, 2017.

\bibitem{2016Perceptual}
Justin Johnson, Alexandre Alahi, and Li~Fei-Fei,
\newblock ``Perceptual losses for real-time style transfer and
  super-resolution,''
\newblock in {\em European conference on computer vision}. Springer, 2016, pp.
  694--711.

\bibitem{simonyan2014very}
Karen Simonyan and Andrew Zisserman,
\newblock ``Very deep convolutional networks for large-scale image
  recognition,''
\newblock {\em arXiv preprint arXiv:1409.1556}, 2014.

\bibitem{PSNRSSIM}
Alain Hore and Djemel Ziou,
\newblock ``Image quality metrics: Psnr vs. ssim,''
\newblock in {\em ICPR}, 2010.

\bibitem{2018FVD}
T.~Unterthiner, S.~van Steenkiste, K.~Kurach, R.~Marinier, M.~Michalski, and
  S.~Gelly,
\newblock ``Towards accurate generative models of video: A new metric \&
  challenges,''
\newblock {\em arXiv preprint arXiv:1812.01717}, 2018.

\bibitem{ji2021audio}
Xinya Ji, Hang Zhou, Kaisiyuan Wang, Wayne Wu, Chen~Change Loy, Xun Cao, and
  Feng Xu,
\newblock ``Audio-driven emotional video portraits,''
\newblock in {\em CVPR}, 2021.

\end{thebibliography}

\end{document}